\documentclass[bibyear]{aa}  
\usepackage{graphicx}
\usepackage{txfonts}
\usepackage{lscape}

\newenvironment{itemlist}{
\begin{list}{}{\setlength{\leftmargin}{10mm}\setlength{\parsep}{0mm}
\setlength{\itemsep}{3mm}} }{\end{list}}
\usepackage{color}

\begin{document}

   \title{Large Halloween Asteroid at Lunar Distance\thanks{Analysis is also based
      on observations collected at the European Southern Observatory, Chile;
      ESO, DDT proposal 296.C-5007(A)}$^{,}$ \thanks{Our visual lightcurve observations are only available in electronic form
          at the CDS via anonymous ftp to cdsarc.u-strasbg.fr (130.79.128.5) or via
          http://cdsweb.u-strasbg.fr/cgi-bin/qcat?J/A+A/}}


   \author{
          T.\ G.\ M\"{u}ller \inst{1},
          A.\ Marciniak \inst{2},
          M.\ Butkiewicz-B\k{a}k \inst{2},
          R.\ Duffard \inst{3},
          D.\ Oszkiewicz \inst{2}
          H.\ U.\ K\"aufl \inst{4},
          R.\ Szak{\'a}ts \inst{5},
          T.\ Santana-Ros \inst{2},
          C.\ Kiss \inst{5},
          \and
          P.\ Santos-Sanz \inst{3}
          }

   \institute{   {Max-Planck-Institut f\"{u}r extraterrestrische Physik,
                 Postfach 1312, Giessenbachstra{\ss}e,
                 85741 Garching, Germany
                 }
          \and   {Astronomical Observatory Institute, Faculty of Physics,
                 A.\ Mickiewicz University, S{\l}oneczna 36, 60-286 Pozna{\'n}, Poland
                 }
          \and   {Instituto de Astrof\'isica de Andaluc\'ia (CSIC)
                 C/ Camino Bajo de Hu\'etor, 50, 18008 Granada, Spain
                 }
          \and   {ESO, Karl-Schwarzschild-Str.\ 2,
                 85748 Garching, Germany
                 }
          \and
                 {Konkoly Observatory, Research Center for Astronomy and
                 Earth Sciences, Hungarian Academy of Sciences;
                 Konkoly Thege 15-17, H-1121 Budapest, Hungary
                 }
          }

   \date{Received ; accepted }

\abstract{The near-Earth asteroid (NEA) \object{2015~TB$_{145}$} had a very close encounter with
          Earth at 1.3 lunar distances on October 31, 2015. We obtained 3-band mid-infrared
          observations of this asteroid with the ESO VLT-VISIR instrument covering approximately
          four hours in total. We also monitored the visual lightcurve during the close-encounter phase.
          The NEA has a (most likely) rotation period of 2.939 $\pm$ 0.005 hours and
          the visual lightcurve shows a peak-to-peak amplitude of approximately 0.12 $\pm$ 0.02\,mag.
          A second rotation period of 4.779 $\pm$ 0.012\,h, with an amplitude of
          the Fourier fit of 0.10 $\pm$ 0.02\,mag, also seems compatible with the available
          lightcurve measurements.
          We estimate a V-R colour of 0.56 $\pm$ 0.05\,mag from different entries in the MPC database.
          A reliable determination of the object's absolute magnitude was not possible.
          Applying different phase relations to the available R-/V-band observations
          produced H$_R$ = 18.6\,mag (standard H-G calculations) or 
          H$_R$ = 19.2\,mag \& H$_V$ = 19.8\,mag (via the H-G$_{12}$ procedure for sparse and
          low-quality data), with large uncertainties of approximately 1\,mag.
          We performed a detailed thermophysical model analysis by using spherical and partially
          also ellipsoidal shape models. The thermal properties are best explained by an 
          equator-on ($\pm$ $\approx$30$^{\circ}$) viewing geometry during our measurements
          with a thermal inertia in the range 250-700\,Jm$^{-2}$s$^{-0.5}$K$^{-1}$ (retrograde rotation)
          or above 500\,Jm$^{-2}$s$^{-0.5}$K$^{-1}$ (prograde rotation). We find that the NEA has
          a minimum size of approximately 625\,m, a maximum size of just below 700\,m, and a slightly
          elongated shape with a/b $\approx$1.1. The best match to all thermal measurements is
          found for: (i) Thermal inertia $\Gamma$ = 900\,Jm$^{-2}$s$^{-0.5}$K$^{-1}$; D$_{eff}$ = 644\,m,
          p$_V$ = 5.5\% (prograde rotation with 2.939\,h); regolith grain sizes of $\approx$50-100\,mm;
          (ii) thermal inertia $\Gamma$ = 400\,Jm$^{-2}$s$^{-0.5}$K$^{-1}$; D$_{eff}$ = 667\,m,
          p$_V$ = 5.1\% (retrograde rotation with 2.939\,h); regolith grain sizes of $\approx$10-20\,mm.
          A near-Earth asteroid model (NEATM) confirms an object size well above 600\,m (best NEATM
          solution at 690\,m, beaming parameter $\eta$ = 1.95), significantly larger than early estimates
          based on radar measurements.
          In general, a high-quality physical and thermal characterisation of a close-encounter object
          from two-week apparition data is not easily possible. We give recommendations for improved
          observing strategies for similar events in the future.
          }


\authorrunning{M\"uller et al.}
\titlerunning{Large Halloween Asteroid at Lunar Distance}

   \maketitle
%

\section{Introduction}

The Apollo-type near-Earth asteroid \object{(NEA) 2015~TB$_{145}$} was discovered by
Pan-STARRS\footnote{Panoramic Survey Telescope \& Rapid Response System: {\tt http://pan-starrs.ifa.hawaii.edu/public/}}
on October 10, 2015. It is on a highly eccentric (e = 0.86) and inclined
(i = 39.7$^{\circ}$) orbit with a semi-major axis of 2.11\,AU, a perihelion
distance of 0.29\,AU and an aphelion at 3.93\,AU. Its current minimum orbital intersection
distance (MOID) with Earth is at 0.0019\,AU.
The pre-encounter H-magnitude estimate of 19.8\,mag indicated a
size range of 200\,m (assuming a high albedo of 50\%) up to 840\,m (assuming a very
dark surface with 3\% albedo). For comparison: Apophis, the object with the currently
highest impact risk on the Torino scale, has a size of approximately 375\,m in diameter and a
high albedo of 30\% (M\"uller et al.\ \cite{mueller14}; Licandro et al.\
\cite{licandro16}).
Based on its MOID and H-magnitude estimate, it is considered as a Potentially
Hazardous Asteroid (PHA: H $<$ 22.0\,mag and MOID $\le$ 0.05\,AU).
On Oct.\ 31, 2015 it passed Earth at approximately 1.3 lunar distances.
It was the closest approach of an object of that size since 2006, the
next (known) similar event is the passage of 137108 (1999 AN$_{10}$) on Aug.\ 7, 2027.
99942~Apophis will follow on Apr.\ 13, 2029 with an Earth passage at approximately 0.1 lunar distances.
 
The close Earth approach made 2015~TB$_{145}$ an important
reference target for testing various techniques to characterise the object's
properties, required for a long-term orbit prediction based on gravitational
and non-gravitational forces. There were several ongoing,  ground-based observing campaigns
to obtain simultaneous visual lightcurves. NASA took advantage of this truly outstanding
opportunity to obtain radar images with 2\,m/pixel resolution via
Green Bank, and Arecibo antennas\footnote{Goldstone Radar Observations Planning:
2009~FD and 2015~TB$_{145}$. NASA/JPL Asteroid Radar Research. Retrieved
2015-10-22: {\tt http://echo.jpl.nasa.gov/\-asteroids/\-2009FD/\-2009FD\_planning.html}}.
The lightcurve measurements, in combination with radar data, will help to
characterise the object's shape and rotation period.
The thermal measurements with VISIR\footnote{The VLT spectrometer and imager for the mid-infrared VISIR}
are crucial for deriving the object's size and albedo via radiometric techniques (see e.g. Delbo et
al. \cite{delbo15}, and references therein). The multi-wavelength coverage of the thermal N-band
lightcurve will also contain information about the object's cross-section, but more importantly, it
allows us to constrain thermal properties of the object's surface (e.g. M\"uller et al.\ \cite{mueller05}).
The derived properties such as size, albedo, shape, spin properties and thermal inertia  can then be used
for long-term orbit calculations and impact risk studies which require careful consideration
of the Yarkovsky effect, a small, but significant non-gravitational force
(Vokrouhlicky et al.\ \cite{vokrouhlicky15}).

In this paper, we first present our ESO-VISIR Director Discretionary awarded
Time (DDT) observations of 2015~TB$_{145}$, including the data reduction
and calibration steps.
The results from lightcurve observations, absolute measurements
and colours are described in Section~\ref{sec:visual}.
We follow this with a radiometric analysis using all available data
by means of our thermal model and present the derived
properties (Section \ref{sec:tpm}). In Section~\ref{sec:dis} we
discuss the object's size, shape and albedo in a wider context and
study the influences of surface roughness, thermal inertia, rotation period
and H-magnitude in more detail. We finally conclude the paper with a
summary of all derived properties and with the implications for observations
of other Potentially Hazardous Asteroids (PHAs).

\section{Mid-infrared observations with ESO VLT-VISIR}
\label{sec:obs}

We were awarded DDT to observe 2015~TB$_{145}$ in October 2015
via ground-based N-band observations with the ESO-VISIR instrument
(Lagage et al.\ \cite{lagage04}) mounted on the 8.2\,m VLT telescope
MELIPAL (UT~3) on Paranal. The work presented here is, to the best of
our knowledge, the first publication after the upgrade of VISIR (K\"aufl et al.\ \cite{kaeufl15}).

The service-mode observers worked very hard to execute our observing
blocks (OB) in the only possible observing window on October 30, 2015
(stored in ESO archive under 2015-10-29).
All OBs were done in imaging mode, each time including  the J8.9
($\lambda_c$ = 8.72\,$\mu$m), SIV\_2 (10.77\,$\mu$m), and the
PAH2\_2 (11.88\,$\mu$m) filters. The 2015~TB$_{145}$ observations
were performed in parallel nod-chop mode with a throw of 8$^{\prime \prime}$
and a chopper frequency of 4\,Hz, a detector integration time of 0.0125\,s
(ten integrations per chopper half cycle where seven integrations are used
in the data reduction), and a pixel field-of-view of 0.0453$^{\prime \prime}$.

\begin{table*}[h!tb]
  \caption{Observation log summary. The ESO VLT-VISIR observations of 
           the calibration star HD26967 are related to the programme ID
           60.A-9234(A), while the HD37160 and our science target observations
           are related to the programme ID 296.C-5007(A). The observing
           date (UT) was 2015-Oct-30.
     \label{tbl:obssequence}}
{\tiny
     \begin{tabular}{lllll}
        \noalign{\smallskip}
        \hline
        \hline
        \noalign{\smallskip}
UT [hh:mm] & source  & filter sequence   & AM range  & remarks \\
        \noalign{\smallskip}
        \hline
        \noalign{\smallskip}
05:05 - 05:15 & HD26967 & 1$\times$ (J8.9, SIV\_2, PAH2\_2) & 1.09...1.08 & standard calibration 60.A-9234(A) \\
05:30 - 05:40 & NEA     & J8.9                              & 1.18...1.16 & target acquisition \\
05:44 - 07:58 & NEA     & 7$\times$ (J8.9, SIV\_2, PAH2\_2) & 1.12...1.19 & lightcurve sequence 1 \\
08:01 - 08:17 & HD37160 & 1$\times$ (SIV\_2, PAH2\_2, J8.9) & 1.21...1.22 & standard calibration 296.C-5007(A) \\
08:21 - 08:24 & NEA     & J8.9                              & 1.24...1.25 & target acquisition \\
08:25 - 09:05 & NEA     & 2$\times$ (J8.9, SIV\_2, PAH2\_2) & 1.26...1.39 & lightcurve sequence 2 \\
09:06 - 09:10 & NEA     & J8.9                              & 1.41...1.42 & lightcurve sequence 2 (con't) \\
09:14 - 09:29 & HD37160 & 1$\times$ (SIV\_2, PAH2\_2, J8.9) & 1.31...1.35 & standard calibration 296.C-5007(A) \\
     \end{tabular}
}
\end{table*}

Table~\ref{tbl:obssequence} shows a summary of the observing logs. In each band
we executed on-array chopped measurements at nod positions A, B, A, and B again.
For the calibration stars, the data have been stored in a nominal way for each
full nod cycle. For our extremely-fast moving NEA, the data had to be recorded in
half nod cycles, that is,\ one image every 0.125\,s, to avoid elongated Point-spread functions (PSFs)
due to the significant field-rotation in the auto-guiding system.
Due to the frequent manual interventions of the operator,
not all OBs could be executed during the proposal-related 3.5\,h schedule slot.
The science blocks were taken in tracking mode based on Cerro Paranal centric
JPL Horizons\footnote{\tt http://ssd.jpl.nasa.gov/horizons.cgi} ephemeris
predictions from Oct.\ 29, 2015.

The pipeline-processed images of either half or full nod cycles were
used for aperture photometry. The aperture size was selected separately for each
band  (but identical for calibrators and NEA) with the aim to (i) include
the entire object flux even in cases of elongated or
distorted PSF structures, (ii) to optimise S/N ratios, and (iii) to avoid
background structures or detector artifacts. Aperture photometry was
performed on the positive and negative beams separately.

Table~\ref{tbl:calibration} shows the conversion factors derived from the
star measurements listed in Table~\ref{tbl:obssequence}. The observing
conditions during the 4.5\,h of measurements were very stable and the
average counts-to-Jansky conversion factors changed by only 1-2\% in a given band.
The stars and the NEA were observed at similar airmass (AM) and no correction 
was needed (see Sch\"utz \& Sterzik \cite{schuetz05}).

\begin{table*}[h!tb]
  \caption{Calibration star model fluxes and count-to-Jansky conversion factors.
           The model fluxes are taken from stellar model templates (Cohen et al.\
           \cite{cohen99}) and interpolated to the band reference wavelength.
     \label{tbl:calibration}}
     \begin{tabular}{lrrr}
        \noalign{\smallskip}
        \hline
        \hline
        \noalign{\smallskip}
Filter  & J8.9               & SIV\_2              & PAH2\_2             \\
FD at   & 8.72\,$\mu$m       & 10.77\,$\mu$m       & 11.88\,$\mu$m \\
        & [Jy]               & [Jy]                & [Jy]                \\
        \noalign{\smallskip}
        \hline
        \noalign{\smallskip}
HD26967 & 15.523 & 11.011 & 9.106 \\
HD37160 & 11.470 &  7.704 & 6.374 \\
        \noalign{\smallskip}
        \hline
        \noalign{\smallskip}
\multicolumn{4}{l}{Conversion factors [Jy/s/counts]:} \\
        \noalign{\smallskip}
        \hline
        \noalign{\smallskip}
aper.\ radius   &          30 pixel &          25 pixel &          25 pixel \\
sky annulus     &       50-60 pixel &       50-60 pixel &       50-60 pixel \\
conversion      & 4214.0 $\pm$ 84.9 & 1628.3 $\pm$ 35.8 & 2695.2 $\pm$ 20.7 \\
        \noalign{\smallskip}
        \hline
        \noalign{\smallskip}
\multicolumn{4}{l}{Colour-correction terms: FD$_{\lambda_{c}}$ = FD$_{obs}$/cc\_corr} \\
        \noalign{\smallskip}
        \hline
        \noalign{\smallskip}
cc\_corr & 0.99 & 1.01 & 0.97 \\  
        \noalign{\smallskip}
        \hline
        \noalign{\smallskip}
     \end{tabular}
\end{table*}

\begin{figure}[h!tb]
 \rotatebox{90}{\resizebox{!}{\hsize}{\includegraphics{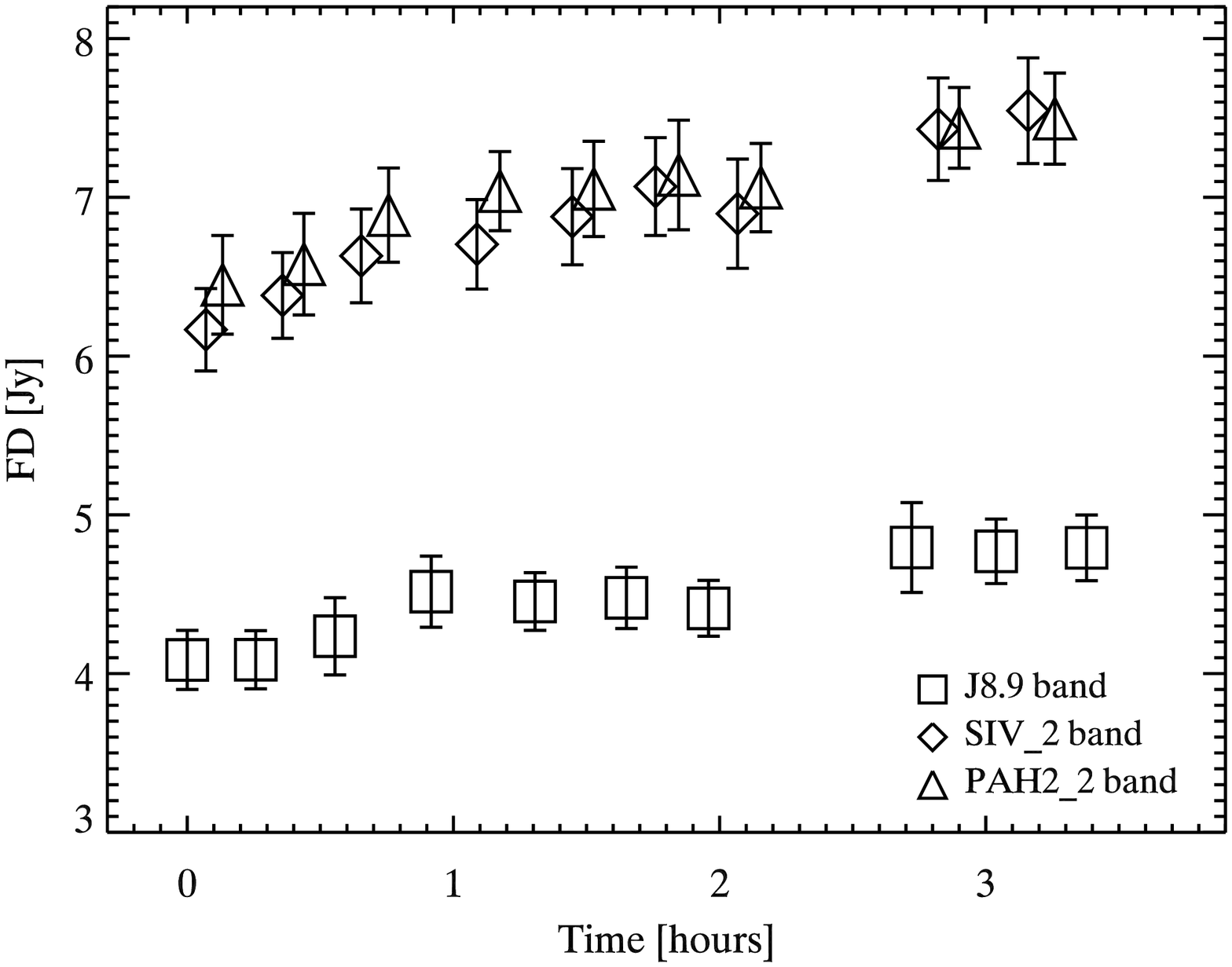}}}
 \rotatebox{90}{\resizebox{!}{\hsize}{\includegraphics{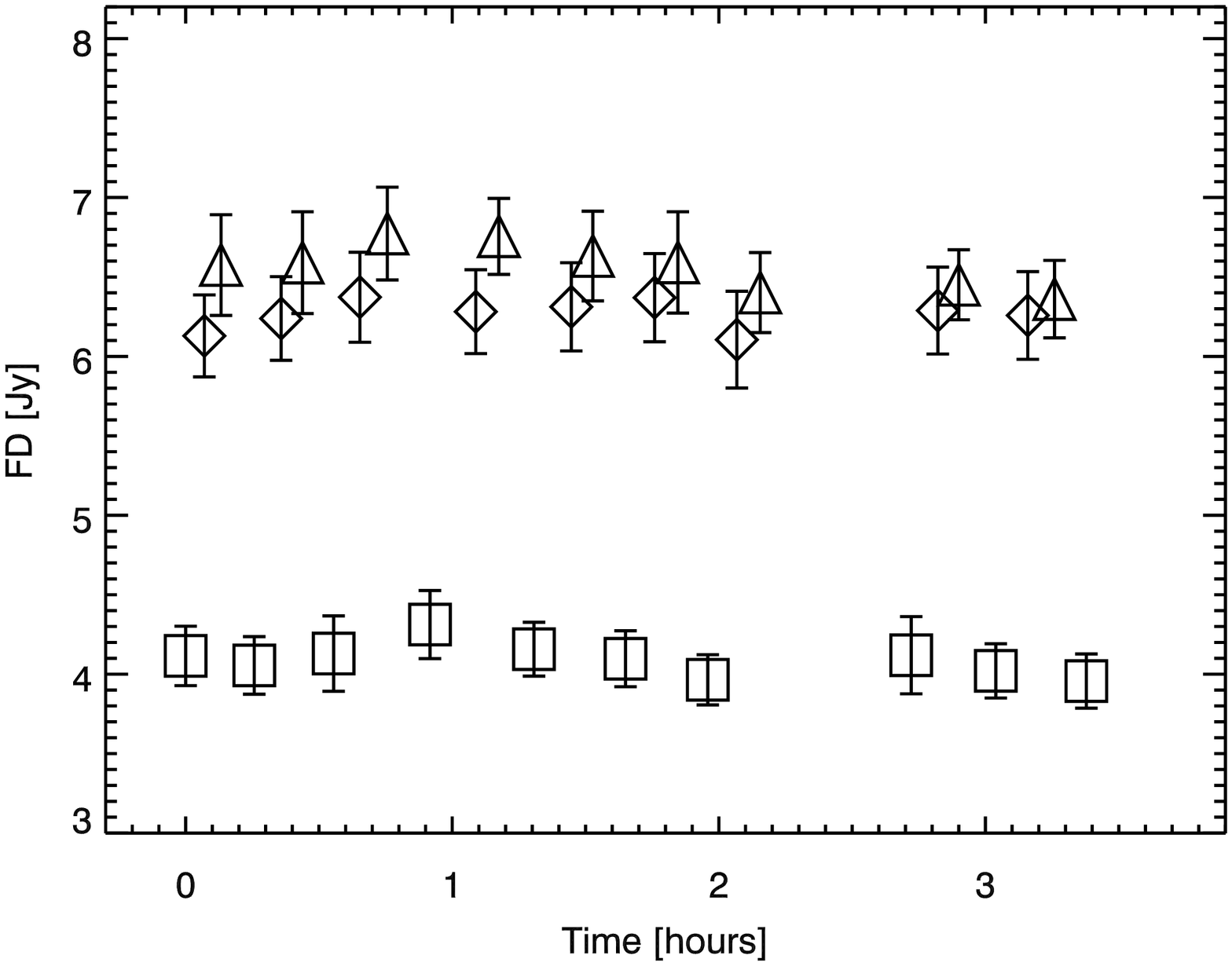}}}
  \caption{The observed and calibrated VISIR flux densities of 2015 TB$_{145}$.
           Zero-time corresponds to Oct.\ 30, 2015 at 05:46 UT.
           Top: The flux increase with time is for the most part related to the rapidly
           changing observing geometry. Bottom: Same fluxes, but now distance
           corrected to the first measurement in a given filter.
     \label{fig:visir1}}
\end{figure}

\begin{table*}[h!tb]
  \caption{Observing geometry for 2015~TB$_{145}$: r$_{helio}$ is the heliocentric distance,
           $\Delta$ is the observatory-object distance, $\alpha$ is the phase angle,
           $\delta$RA$\cdot$cos(DEC) and $\delta$(DEC)/dt are rate of change of target
           center apparent right ascension and declination.
           The observing date (UT) was 2015-Oct-30.
     \label{tbl:obsgeometry}}
     \begin{tabular}{lllllll}
        \noalign{\smallskip}
        \hline
        \hline
        \noalign{\smallskip}
 UT      & r$_{helio}$ & $\Delta$ & $\alpha$     & $\delta$RA$\cdot$cos(DEC) &  $\delta$(DEC)/dt       & \\
 $[$hh:mm] & [AU]        & [AU]     & [$^{\circ}$] & [$^{\prime \prime}$/h]    &  [$^{\prime \prime}$/h] & remarks \\
        \noalign{\smallskip}
        \hline
        \noalign{\smallskip}
05:30 & 1.01777 & 0.02998 & +34.37 & 293.9 & 519.9 & start target acquisition 1 \\
05:40 & 1.01765 & 0.02984 & +34.38 & 295.8 & 524.7 & end target acquisition 1 \\
05:44 & 1.01761 & 0.02979 & +34.38 & 296.6 & 526.6 & start lightcurve sequence 1 \\
07:58 & 1.01604 & 0.02792 & +34.42 & 339.1 & 597.4 & end lightcurve sequence 1 \\
08:21 & 1.01577 & 0.02760 & +34.43 & 349.8 & 610.9 & start target acquisition 2 \\
08:24 & 1.01573 & 0.02756 & +34.43 & 351.3 & 612.7 & end target acquisition 2 \\
08:25 & 1.01572 & 0.02755 & +34.43 & 351.8 & 613.3 & start lightcurve sequence 2 \\
09:10 & 1.01519 & 0.02693 & +34.45 & 376.2 & 641.4 & end lightcurve sequence 2 \\
        \noalign{\smallskip}
        \hline
        \noalign{\smallskip}
     \end{tabular}
\end{table*}

 We applied colour corrections of 0.99 (J8.9), 1.01 (SIV\_2),
 and 0.97 (PAH2\_2) to obtain the object's mono-chromatic
 flux density at the corresponding band reference wavelengths (see
 Table~\ref{tbl:calibration}). These corrections are based on
 stellar model SEDs for both stars, our best pro- and retrograde
 model predictions for 2015~TB$_{145}$, and the corresponding VISIR
 filter transmission curves. For the error calculation
 we quadratically added the following error sources:
 error in count-to-Jy conversion factor (2\%), error of stellar
 model (3\%) and aperture photometry error as given by the standard
 deviation of the eight photometric data points of the two full nod
 cycles with two positive and two negative beams each (1...5\%),
 summing up to a total of 4-6\% error
 in the derived absolute flux densities (see Table~\ref{tbl:obsvisir}
 and Fig.~\ref{fig:visir1}).

\section{Photometric observations}
\label{sec:visual}

\subsection{Results from lightcurve observations}
\label{sec:lc}

To obtain the rotational period of 2015~TB$_{145}$ , we planned
several observing runs at different telescopes in Spain: the 1.23-m telescope
at Calar Alto Observatory (CAHA) in Almeria,  the 1.5-m telescope
at Sierra Nevada Observatory (OSN) in Granada, and the 0.80-m telescope
at La Hita Observatory, near Toledo.

The OSN observations were carried out by means of a 2k x 2k CCD\footnote{Charge-Coupled Device (CCD)},
with a total field of view (FOV) of 7.8 x 7.8\,arcmin. We used a 2x2 binning mode, which
provides a scale of image of 0.46\,arcsec/pixel. Observations with this
telescope were made from Oct.\ 29, 2015 23:56\,UT to Oct.\ 30, 2015 04:22\,UT
using a Johnson R-filter, obtaining a total of 930 images with 15\,sec
exposure time each frame.
Due to the object's extremely fast motion, the extraction of useful
information for the lightcurve reconstruction proved very difficult, however it
was possible to obtain calibrated R-band magnitudes of the asteroid using the data.

The CAHA observing run was executed from Oct.\ 30, 2015 22:18\,UT to
Oct.\ 31, 2015 05:59\,UT using the 4k x 4k DLR-MKIII CCD camera of the
1.23-m Calar Alto Observatory telescope. The image scale and the
FOV of the instrument are 0.32\,arcsec/pixel and 21.5 x 21.5\,arcmin,
respectively. The images were obtained in 2x2 binning mode and were taken
using the clear filter. A total of 158 science images were obtained (distributed
over 21 fields of view) with an integration time of 1 second.
Bias frames and twilight sky flat-field frames
were taken each night for the three telescopes. These bias and flatfields
were used to properly calibrate the images. Relative photometry was obtained
for each night and each FOV using as many stars as possible to minimise
errors in photometry. These data are used in the Fourier analysis
to find the true rotation period.

The La Hita Observatory observations were carried out by means of a
4k x 4k CCD, with a FOV of 47 x 47\,arcmin. Observations with this
telescope were made from Oct.\ 30, 2015 03:12 to 05:36\,UT using no
filter to reach larger signal-to-noise ratios (S/N), and obtaining
a total of 144 images with 15\,sec exposure time each one. A second night
at La Hita telescope produced a total of 147 images with 5\,sec exposure
time for each image from Oct.\ 30, 2015 21:15 to 23:19\,UT, also without filter.
Due to the fast apparent movement of the object on the sky plane, two
different FOVs were acquired in the second night, therefore different reference
stars were used to do the relative photometry. However, the final quality
of these measurements was not sufficient to constrain the object's
rotation period.

Another lightcurve from Chile (small telescope BEST at Cerro Armazones) taken
on Oct.\ 31, 2015 also had to be excluded because this fragment fitted neither in
amplitude nor any period. The data were taken at similar phase angle as our
other lightcurves, however instrumental effects due to the object's
fast motion, proximity to the Moon, and smearing are very likely to have had a severe impact on the
data quality.

In addition to our lightcurves, we used another three
datasets from the observer J.\ Oey (Australia)
available from the MPC lightcurve database (data from Oct.\ 24, 29, and 30 (another
set from Oct.\ 25 was too noisy and was therefore excluded)), and a large collection of
data presented in Warner et al.\ (\cite{warner16}) and provided by B.\ Warner
(priv.\ communication, Jul.\ 2016).

\begin{figure}[h!tb]
 \rotatebox{0}{\resizebox{\hsize}{!}{\includegraphics{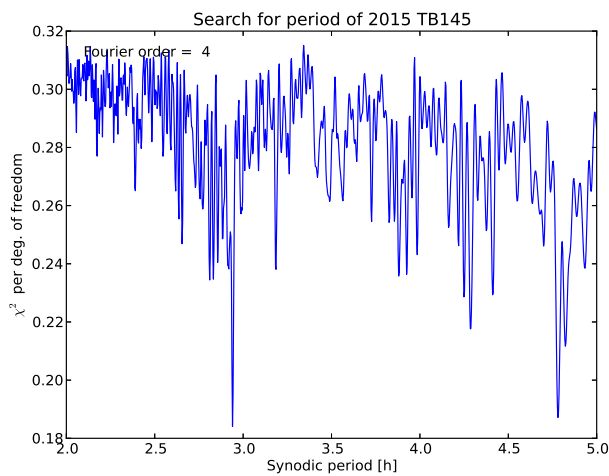}}}
  \caption{The resulting periodogram from our lightcurve analysis. There are
           two possible solutions: at (synodic) rotation period of 2.939\,h (lowest
           $\chi^2$ solution), and 4.779\,h (second best solution). Others
           have more than 10\% higher $\chi^2$ values.
     \label{fig:periodogram}}
\end{figure}

\begin{figure}[h!tb]
 \rotatebox{0}{\resizebox{\hsize}{!}{\includegraphics{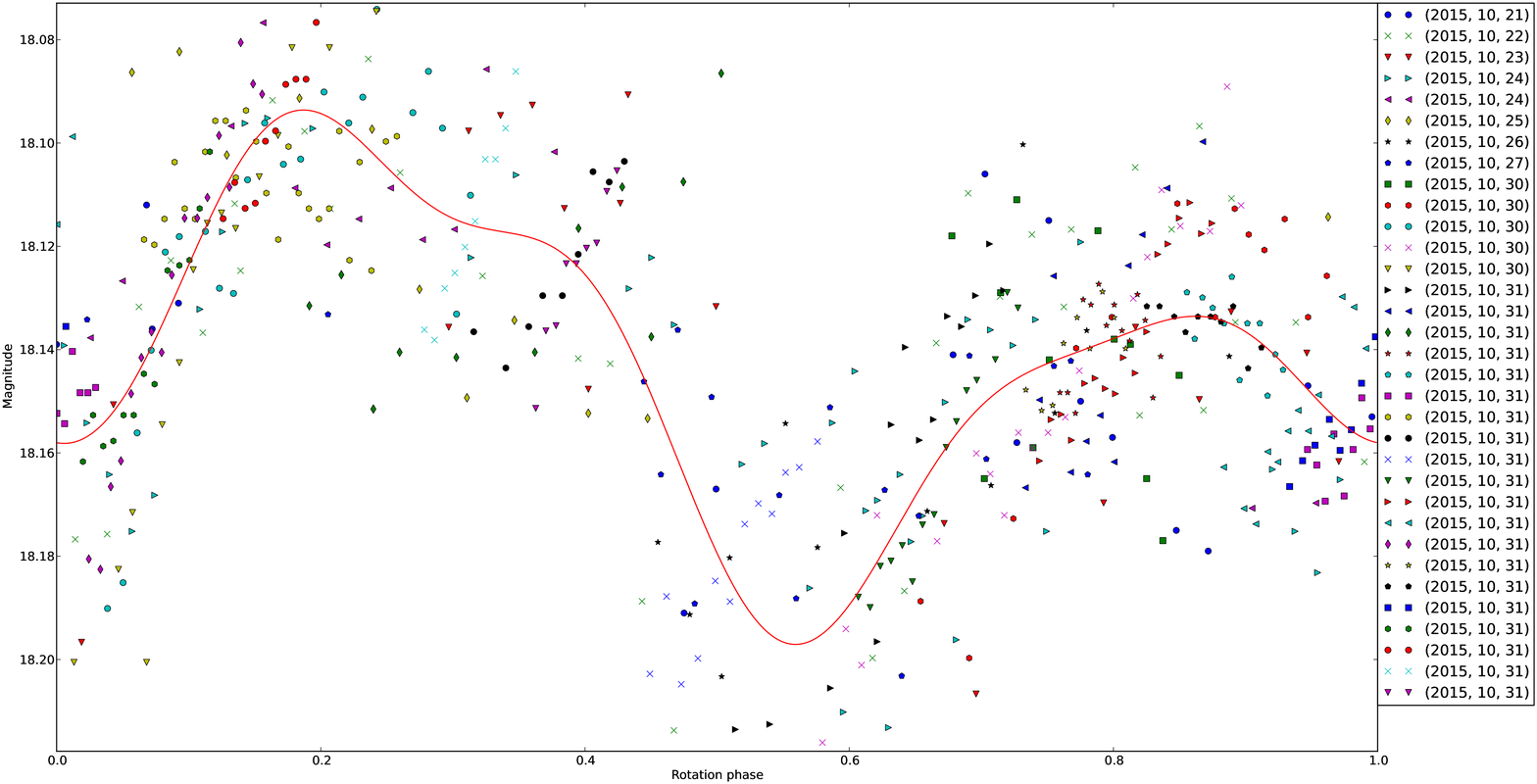}}}
  \caption{The best composite lightcurve from our Fourier analysis
           using relative magnitudes. Different telescopes
           and dates are indicated with different colours and/or symbols. 
           The rotation period is 2.939 $\pm$ 0.005\,h, with an amplitude of
           the 4$^{th}$-order Fourier fit of 0.12 $\pm$ 0.02\,mag. The zero date used
           to fold the data is 2015 Oct.\ 21.3771 UTC (LT corrected).
     \label{fig:best_period_lc}}
\end{figure}

\begin{figure}[h!tb]
 \rotatebox{0}{\resizebox{\hsize}{!}{\includegraphics{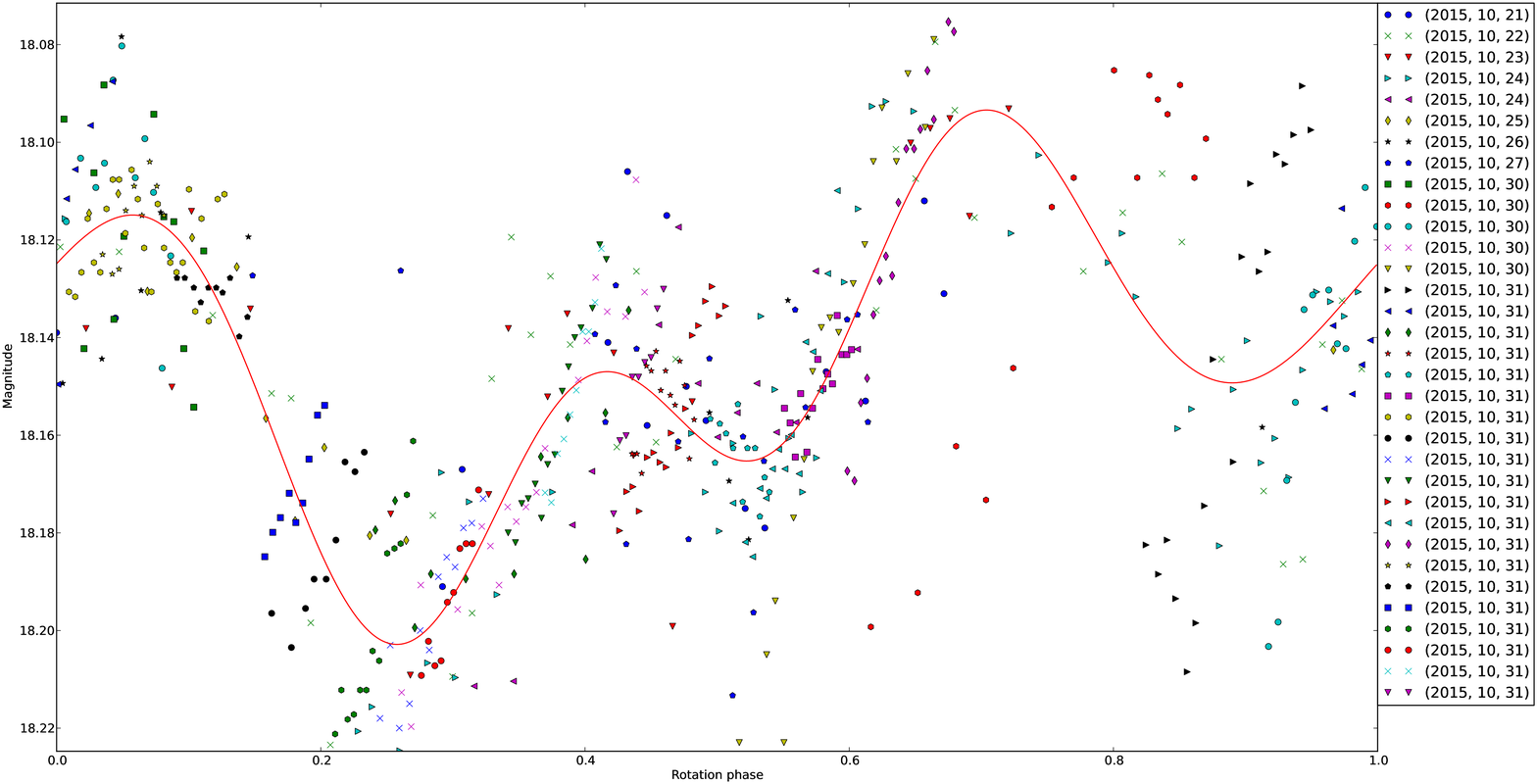}}}
  \caption{The second best composite lightcurve from our Fourier analysis.
           The rotation period is 4.779 $\pm$ 0.012\,h, with an amplitude of
           the 4$^{th}$-order Fourier fit of 0.10 $\pm$ 0.02\,mag. The zero date used
           to fold the data is 2015 Oct.\ 21.3771 UTC (LT corrected).
     \label{fig:secondbest_period_lc}}
\end{figure}

\begin{figure}[h!tb]
 \rotatebox{0}{\resizebox{\hsize}{!}{\includegraphics{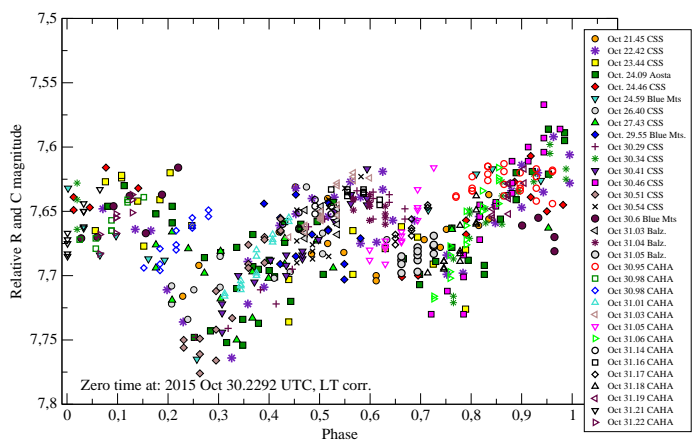}}}
  \caption{The composit lightcurve of 2015~TB$_{145}$ with a zero phase
           close to the start of the VLT-VISIR measurements
           (based on P = 2.941 $\pm$ 0.005\,h, amplitude
           of 0.15 $\pm$ 0.03\,mag, and zero phase at 2015, Oct 30.2292 UTC,
           LT corr.).
     \label{fig:composite}}
\end{figure}

On this combined dataset, we performed a Fourier
analysis (see periodogram in Fig.~\ref{fig:periodogram})
and found the best period at 2.939 $\pm$ 0.005\,hours
with the model lightcurve having an amplitude of 0.12 $\pm$ 0.02\,mag
(see Fig.~\ref{fig:best_period_lc}), very close to the Warner et al.\
(\cite{warner16}) solution with 2.938 $\pm$ 0.002\,h rotation period
and 0.13 $\pm$ 0.02\,mag amplitude. The second-lowest $\chi^2$ is found
for a rotation period at 4.779\,h (see Fig.~\ref{fig:secondbest_period_lc}).
We consider the second solution less probable, as it implies a more
complicated rarely-occurring 3-maxima lightcurve.
Other solutions have $\chi^2$ values that are more than 10\% higher
than the one for the best solution and lead to a severe misfit of various
fragments relative to one another. They can be excluded with high probability.
To verify our results, we also used another algorithm based on pure $\chi^2$
fitting to search for the best composite lightcurves within a given range.
The resulting composite
is shown in Fig.~\ref{fig:composite}, with a start date close to our first
VLT-VISIR measurement. Both methods find the same (best) solution within the
given error bars, given the poor quality of the data and large
uncertainty range.

\subsection{Absolute V-/R-magnitudes}
\label{sec:absmag}
 
Absolute magnitudes were obtained using the images from the OSN telescope as they
were the only images obtained with a filter; the R-Johnson filter. For each star
in the FOV with R magnitude in the USNOB1 catalog, we determined the magnitudes of the NEA and
the star in 3-5 different images. The error was assumed to be the dispersion of the measured
star magnitudes with respect to the catalogue star magnitudes. The obtained asteroid
magnitude was then corrected by geocentric and heliocentric distances and by phase
angle. Following the description in Bowell et al.\ (\cite{bowell89}),
we calculated the H$_R$ magnitude:\\

\begin{center}
\begin{equation}
H_R (1, 1, \alpha) = R_{mag} - 5.0 \cdot log(r \cdot \Delta)
\end{equation}
\end{center}

\begin{center}
\begin{equation}
H_R = H_R (1, 1, \alpha) + 2.5 \cdot log((1-G) \cdot \phi~1 + G \cdot \phi~2)
\end{equation}
\end{center}

with $\alpha$ being the phase angle, R$_{mag}$ the calibrated R-band magnitude of
our target in the images, r the heliocentric distance, $\Delta$ the distance to
the observer, H$_R$ (1, 1, $\alpha$) is the R-band magnitude, at solar phase angle $\alpha$
reduced to unit heliocentric and geocentric distance, H$_R$ is the absolute magnitude (at
$\alpha$ = 0$^{\circ}$), G the slope parameter (here we used the default value of 0.15), and
$\phi$1\footnote{$\phi$1 = exp[-3.33 $\cdot$ (tan($\alpha$)$^{1/2}$)$^{0.63}$]}
and $\phi$2\footnote{$\phi$2 = exp[-1.87 $\cdot$ (tan($\alpha$)$^{0.5}$)$^{1.22}$]}
are two specified phase functions that are normalised to unity at $\alpha$ = 0$^{\circ}$.

After the correction, we obtained the median of this H$_R$
for the asteroid. In total, we determined seven different values (three for the night of Oct.\ 30,
and four for the night of Oct.\ 29). The median of all seven 
values is H$_R$ = 18.7 $\pm$ 0.2\,mag.

As an alternative approach we used the Phase Curve Analyser tool\footnote{\tt http://asteroid.astro.helsinki.fi/astphase}
(Oszkiewicz et al.\ \cite{oszkiewicz11}; Oszkiewicz et al.\ \cite{oszkiewicz12}) and entered all
available V- and R-band measurements from the MPC database\footnote{\tt http://www.minorplanetcenter.net/db\_search}.
The phase curve analyser tool does not put any restraints on the 
slope parameters. Both absolute magnitude and slope parameters are fitted 
simultaneously, thus accounting for the different slopes of various 
taxonomic types. We found the following results:
\begin{itemlist}
\item[$\bullet$] Based on the Bowell et al. (\cite{bowell89}) H-G conventions, the phase curve analyser tool produced H$_R$ = 18.6 (with rms of the fit: 0.29\,mag)
                 and H$_V$ = 19.3\,mag (rms: 0.25\,mag). This H$_R$ solution is in excellent agreement with our own calibrated OSN
                 measurements.
\item[$\bullet$] Based on Muinonen et al. (\cite{muinonen10}) H-G$_{12}$ conventions, the phase curve analyser tool
                 produced H$_R$ = 19.18\,mag and H$_V$ = 19.75\,mag.
                 This is significantly different from the standard H-G assumptions, but, according to the authors, the H,G$_{12}$ phase
                 function is applicable to asteroids with sparse or low-accuracy photometric data, which is the case for 2015~TB$_{145}$.
\end{itemlist}

We use the H-G$_{12}$ solution in the following analysis and also
discuss the impact of
different values. It should be mentioned, however, that a reliable H magnitude cannot
be derived from our data nor from the entries in the MPC astrometric database.
The above values are, in effect, covering all possible slope values, which are
not very well constrained. From the fitting of various phase functions (H-G, H-G$_{12}$, H-G$_1$-G$_2$)
and taking the whole range of all possible solutions into account, we conservatively
estimated a H-mag uncertainty of approximately 1\,mag.

\subsection{V-R colour}
\label{sec:colour}

The above analysis leads to V-R colours of 0.7\,mag (in the H-G system)
and 0.6\,mag (in the H-G$_{12}$ system).
We also searched the MPC entries for measurements with V- and
R-band observations taken by the same observatory (obs.\ code) during
the same night (see Table~\ref{tbl:mpcentries}).

\begin{figure}[h!tb]
 \rotatebox{90}{\resizebox{!}{\hsize}{\includegraphics{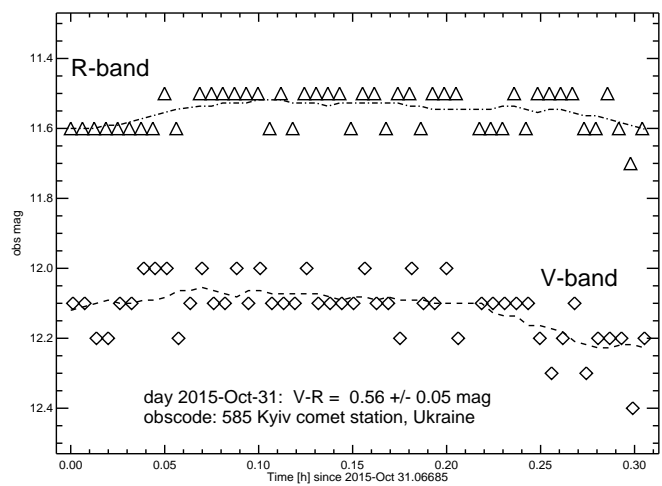}}}
  \caption{V-R colour determination based on selected MPC entries.
     \label{fig:colour}}
\end{figure}


\begin{table}[h!tb]
  \caption{Number of MPC entries where V- and R-band measurements are listed
           for the same night by the same observer (observatory code). All
           measurements are from October 2015.
     \label{tbl:mpcentries}}
\begin{tabular}{llrrl}
\noalign{\smallskip}
\hline\hline
\noalign{\smallskip}
obs. & time span & \multicolumn{2}{c}{meas.\ \#} &   \\
code & [days]      & V  & R              & comments \\
\noalign{\smallskip}
\hline
\noalign{\smallskip}
C48\tablefootmark{1} & 22.778 ... 22.824 & 18 & 23 & interchanged \\ 
Q62\tablefootmark{2} & 24.540 ... 24.612 &  6 &  8 & sequential \\   
Q62                  & 26.503 ... 26.767 & 26 &  3 & sequential \\   
C48                  & 28.759 ... 28.774 &  9 &  4 & interchanged \\ 
C48                  & 29.839 ... 29.842 & 13 &  8 & interchanged \\ 
585\tablefootmark{3} & 31.065 ... 31.080 & 49 & 60 & interchanged \\ 
\noalign{\smallskip}
\hline
\noalign{\smallskip}
\end{tabular}
\tablefoot{\tablefoottext{1}{C48: Sayan Solar Observatory, Irkutsk, Russia};
           \tablefoottext{2}{Q62: iTelescope Observatory, Siding Spring, Australia};
           \tablefoottext{3}{585: Kyiv comet station, Ukraine.}}
\end{table}

The entries from the Kyiv comet station (Ukraine) from Oct.\ 31, 2015, are the most consistent,
and V- and R-band measurements are alternating. The summary
of all data is shown in Fig.~\ref{fig:colour}. Based on these data we calculated a
V-R = 0.56 $\pm$ 0.05\,mag which is confirmed by data from Oct.\ 22, 2015
for obscode C48 (0.55\,mag), and also from Oct.\ 23 for obscode Q62 (0.57\,mag).
The other nights or datasets are more problematic with V- and R-band data appearing poorly
balanced, covering different time periods or with large outliers.
Our derived V-R finding agrees with values documented for other
NEAs such as those by Pravec et al.\ (\cite{pravec95}) or Lin et al.\ (\cite{lin14}), for example.

\section{Radiometric analysis}
\label{sec:tpm}

\begin{figure}[h!tb]
 \rotatebox{90}{\resizebox{!}{\hsize}{\includegraphics{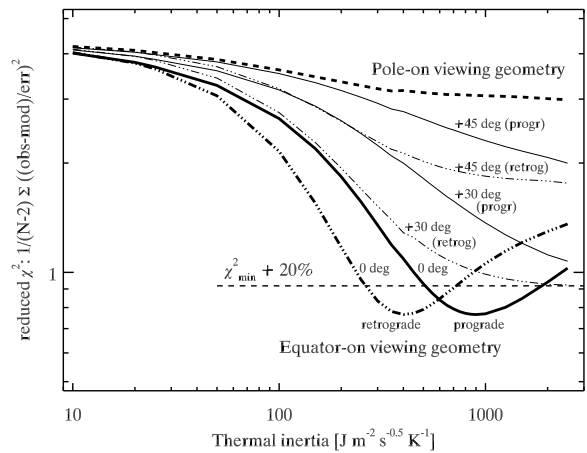}}}
  \caption{The $\chi^2$ radiometric analysis of the VLT-VISIR data for a spherical
           shape solution, and for a range of different spin-axis orientations.
           The dashed horizontal line indicates a $\chi^2$ threshold 20\% above the
           minium value. The solid lines show prograde cases (0$^{\circ}$, +30$^{\circ}$,
           and +45$^{\circ}$ from a perfect equator-on geometry), the dashed-dotted lines 
           are calculated for the same spin-axis orientations, but for retrograde cases.
           The dashed curve with $\chi^2$-values above 3 shows the pole-on geometry.
     \label{fig:chi2}}
\end{figure}

\begin{figure}[h!tb]
 \rotatebox{90}{\resizebox{!}{\hsize}{\includegraphics{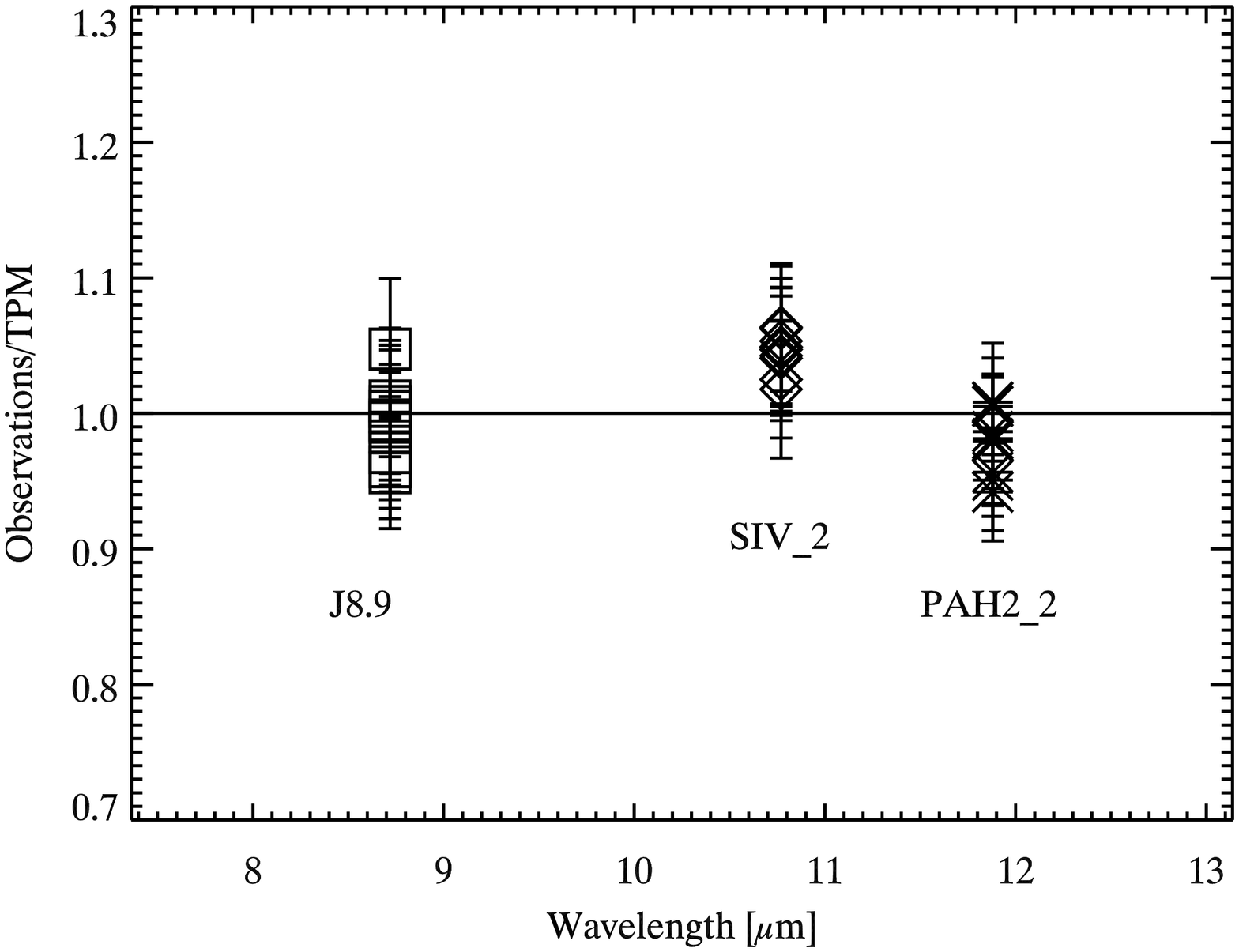}}}
 \rotatebox{90}{\resizebox{!}{\hsize}{\includegraphics{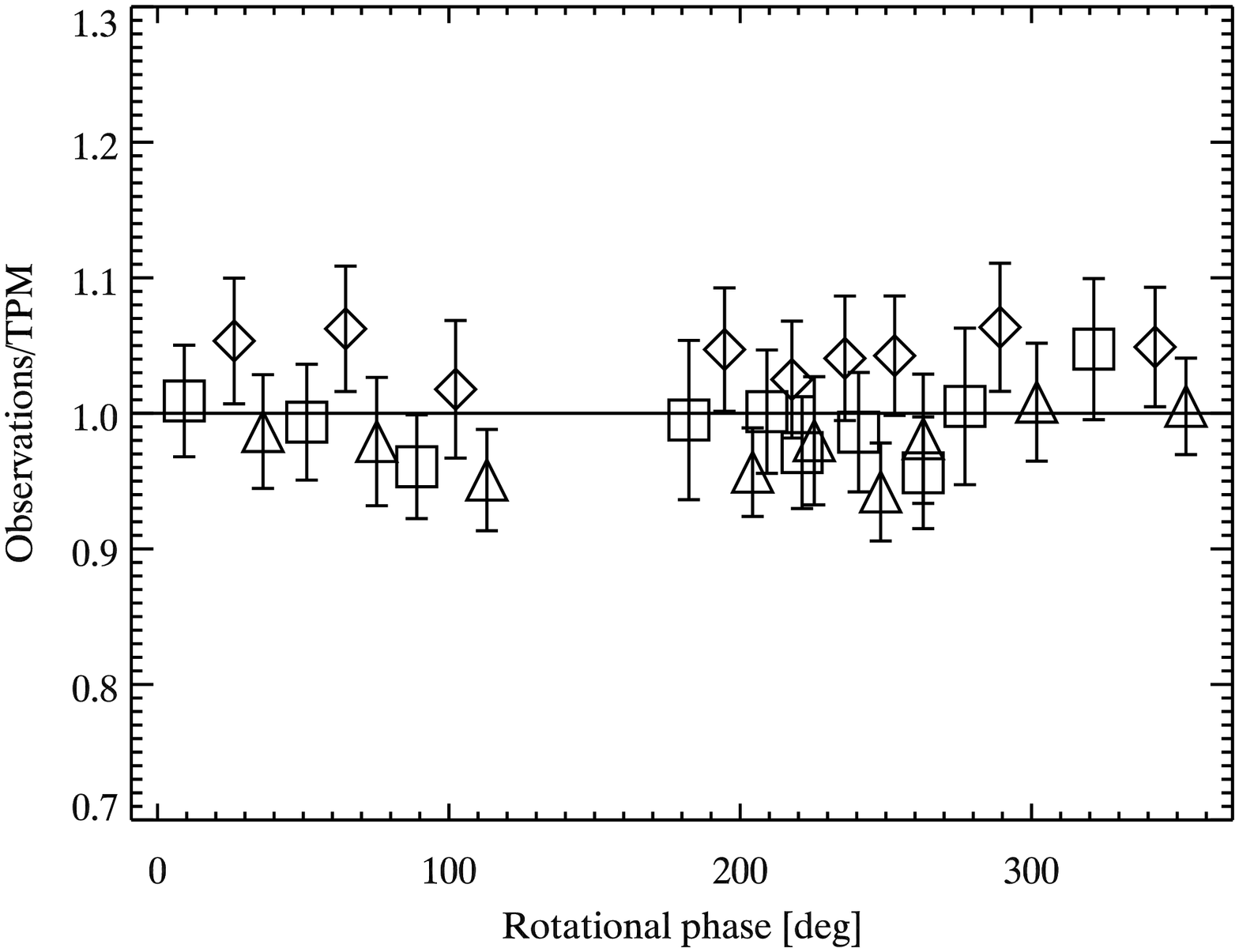}}}
  \caption{The VLT-VISIR observations divided by the corresponding TPM solution,
           assuming a spherical shape with a retrograde rotation (2.939\,h rotation period), a size of 667\,m,
           geometric albedo of 5.1\%, thermal inertia of 400\,Jm$^{-2}$s$^{-0.5}$K$^{-1}$ and
           low surface rougness (r.m.s.\ of surface slopes of 0.1). Top: ratios as a function
           of wavelength; bottom: as a function of rotational phase.
     \label{fig:obsmod}}
\end{figure}

We used the derived properties from Section~\ref{sec:visual} together with
the thermal measurements (Sect.~\ref{sec:obs}) to calculate radiometric
properties of the Halloween asteroid with standard thermophysical model
techniques (see e.g. M\"uller et al.\ \cite{mueller13}
or M\"uller et al.\ \cite{mueller16}). First, we consider a spherical shape
model with a rotation period of 2.939\,h and spin-axis orientations ranging
from pole-on to equator-on during the VLT-VISIR observations. Size, albedo
and thermal properties (thermal inertia $\Gamma$ and surface roughness) are
free parameters in the analysis. Figure~\ref{fig:chi2} shows the $\chi^2$
values for a very wide range of thermal inertias and as a function of
spin-axis orientation. Here, we assumed a low surface roughness
(r.m.s.\ of surface slopes of 0.1). The best fit to the VISIR data
(accepting $\chi^2$ up to 20\% above the minimum $\chi^2$) is
found for thermal inertias in the range 250 to 700\,Jm$^{-2}$s$^{-0.5}$K$^{-1}$
(retrograde rotation) and thermal inertias larger than 500\,Jm$^{-2}$s$^{-0.5}$K$^{-1}$ (prograde rotation).
The lowest $\chi^2$ values are connected to viewing geometries close to equator-on
($\pm$30$^{\circ}$), with the spin axis roughly perpendicular to the line-of-sight.
The best radiometric solutions for an equator-on viewing geometry and a surface roughness
with r.m.s.\ of surface slopes of 0.1 are:
\begin{itemlist}
\item[$\bullet$] prograde rotation (2.939\,h): Thermal inertia $\Gamma$ = 900\,Jm$^{-2}$s$^{-0.5}$K$^{-1}$;
                 D$_{eff}$ = 644\,m, p$_V$ = 5.5\%;
\item[$\bullet$] retrograde rotation (2.939\,h): Thermal inertia $\Gamma$ = 400\,Jm$^{-2}$s$^{-0.5}$K$^{-1}$;
                 D$_{eff}$ = 667\,m, p$_V$ = 5.1\% (see also Fig.~\ref{fig:obsmod}).
\end{itemlist}

Based on the retrograde best solution, we produced plots with observation-to-model ratios
as a function of wavelength and rotational phase (see Fig.~\ref{fig:obsmod}). No obvious deviations or
trends were visible. Our prograde best solution produces a similar
match between measurements and model predictions.

The same analysis for a high surface roughness (r.m.s.\ of surface slopes of 0.5)
leads to thermal inertias shifted to higher values ($\Gamma$ $>$ 500\,Jm$^{-2}$s$^{-0.5}$K$^{-1}$
for retrograde, and $\Gamma$ $>$ 1500\,Jm$^{-2}$s$^{-0.5}$K$^{-1}$ for prograde 
rotation), with very similar size-albedo values. With our thermal dataset limited to a small
wavelength range and a single epoch, it is not possible to break the degeneracy between
thermal inertia and surface roughness: high-inertia combined with high surface roughness
fits equally well as a low-inertia, low-roughness case.

The geometric V-band albedo p$_V$ of 5-6\% is tightly connected to our choice for the
H magnitude. Assuming H$_{V}$ = 19.3\,mag (see H-G results in Section~\ref{sec:absmag})
would immediately lead to larger albedos of 8-9\% whilst a larger H magnitude would produce
smaller albedo values.

We also tested the influence of a longer rotation period of 2015~TB$_{145}$. A slower
rotating body (4.779\,h instead of 2.939\,h) would shift the location of the
$\chi^2$ minima in Fig.~\ref{fig:chi2} to larger values: The best solution for a prograde
rotation would be around a thermal inertia of 1250\,Jm$^{-2}$s$^{-0.5}$K$^{-1}$ whilst the
retrograde case would be best explained with an inertia of approximately 500\,Jm$^{-2}$s$^{-0.5}$K$^{-1}$.
Radiometric size and albedo solutions remain very similar.

The VISIR fluxes (mainly J8.9 data) show a sinusoidal change over time (most easily seen in
Fig.~\ref{fig:obsmod} bottom, box symbols). If we interpret these variations as changes in the
cross-section during the object's rotation, we find a maximum effective size of approximately 680\,m
at thermal lightcurve maximum and approximately 650\,m at lightcurve minimum.
A rotating ellipsoidal shape (rotation axis $c$, equator-on viewing) with a/b=1.09 and
b/c=1.0 would explain such a variation. A similar conclusion can be drawn from the observed
visual lightcurve amplitude $\Delta$Mag = 0.12 $\pm$ 0.02\,mag (see Fig.~\ref{fig:best_period_lc}),
which is pointing towards an axis ratio of a/b = 1.12\footnote{$\Delta$Mag = 2.5$\cdot$log(a/b), with $\Delta$Mag = 0.12\,mag},
very similar to what we see in the varying thermal measurements.
Also, the variations in visual brightness and thermal flux seem to be approximately in phase, with
minima and maxima occurring at similar times. A firm statement is not possible, however, due to
the error bars of the thermal measurements and the large scatter in visual brightness.

\begin{table}[h!tb]
     \caption{Observational results. The ESO VLT-VISIR observations are
              related to the programme-ID 296.C-5007(A). The times and
              fluxes are related to the 2-nod averaged photometry.
              The flux errors include the errors of the aperture
              photometry (standard deviation of the eight individual fluxes
              from the 2-nod cycle images), a 3\% error for the uncertainties
              in the models of the two calibration stars, and the error of
              the calibration factors.
     \label{tbl:obsvisir}}
     \begin{tabular}{lrrr}
        \noalign{\smallskip}
        \hline
        \hline
        \noalign{\smallskip}
Julian Date & $\lambda_{ref}$ & FD    & FD$_{err}$  \\
mid-time    & [$\mu$m]        & [Jy]  & [Jy]      \\
        \noalign{\smallskip}
        \hline
        \noalign{\smallskip}
2457325.74034 &   8.72 & 4.115 & 0.187\\  
2457325.75106 &   8.72 & 4.116 & 0.184\\  
2457325.76347 &   8.72 & 4.265 & 0.245\\  
2457325.77852 &   8.72 & 4.548 & 0.226\\  
2457325.79478 &   8.72 & 4.485 & 0.183\\  
2457325.80909 &   8.72 & 4.509 & 0.194\\  
2457325.82194 &   8.72 & 4.442 & 0.177\\  
2457325.85373 &   8.72 & 4.828 & 0.285\\  
2457325.86695 &   8.72 & 4.804 & 0.204\\  
2457325.88110 &   8.72 & 4.826 & 0.208\\  
        \noalign{\smallskip}
2457325.74326 &  10.77 & 6.129 & 0.258 \\  
2457325.75526 &  10.77 & 6.344 & 0.268 \\  
2457325.76757 &  10.77 & 6.592 & 0.293 \\  
2457325.78568 &  10.77 & 6.664 & 0.280 \\  
2457325.80061 &  10.77 & 6.837 & 0.301 \\  
2457325.81363 &  10.77 & 7.026 & 0.306 \\  
2457325.82648 &  10.77 & 6.856 & 0.342 \\  
2457325.85789 &  10.77 & 7.385 & 0.321 \\  
2457325.87195 &  10.77 & 7.501 & 0.331 \\  
        \noalign{\smallskip}
2457325.74587 &  11.88 & 6.575 & 0.317 \\  
2457325.75860 &  11.88 & 6.708 & 0.326 \\  
2457325.77186 &  11.88 & 7.023 & 0.303 \\  
2457325.78929 &  11.88 & 7.177 & 0.254 \\  
2457325.80398 &  11.88 & 7.191 & 0.306 \\  
2457325.81727 &  11.88 & 7.281 & 0.352 \\  
2457325.83013 &  11.88 & 7.199 & 0.283 \\  
2457325.86117 &  11.88 & 7.584 & 0.259 \\  
2457325.87613 &  11.88 & 7.643 & 0.293 \\  
     \end{tabular}
\end{table}

\section{Discussions}
\label{sec:dis}

\subsection{Rotation period}

The Halloween asteroid is an interesting example for demonstrating the
possibilities and limitations of radiometric techniques in cases of
very limited observational data. The lightcurves cover only approximately ten days in total
and the object had a very high apparent motion (up to several arcsec/sec)
on the sky. It was therefore very difficult to extract reliable photometry
from small field-of-view images and constantly changing reference stars.
The construction of a full composite lightcurve from such short fragments
is always very difficult and problematic. Here, the resulting large number
of short and partly noisy lightcurve snippets led to two possible rotation
periods (approximately 2.94\,h and 4.78\,h) in the periodogram (Fig.~\ref{fig:periodogram}).
A visual inspection of various possible periods was still needed, and the two 
remaining solutions were found to produce an acceptable fit of the
various lightcurve fragments relative to each other.
The thermal VISIR measurements also cover more than three hours, but, due to the
low lightcurve amplitude, they do not put any additional constraints on the rotation period.
In the end, the two derived rotation periods remain. Both have similar $\chi^2$ minima in
the periodogram and the minima are within 10\% of each other.
All the periods are only synodic periods, however. Sidereal periods can
only be determined with the help of full spin and shape model solutions,
using data from multiple apparitions and/or a wide range of observing geometries.
Our synodic periods are based on the single short apparition from October 2015,
and they are only precise to within two or three decimals at best.

\subsection{Spin-axis orientation}

Single apparition visible lightcurves contain only very limited information
about the orientation of the spin axis: a flat lightcurve can indicate
either a pole-on viewing geometry (combined with arbitrary shape) or a
spherical shape (with arbitrary orientation of the spin axis). Thermal data,
however, can provide clues about the spin-axis orientation: a measurement of
the shape of the spectral energy distribution contains information about
the surface temperatures and these temperatures depend on the orientation
of the spin axis and the thermal history of surface elements.
Typically, a pole-on geometry produces much higher sub-solar temperatures
(almost independent of the thermal properties of the surface) than a
equator-on geometry where surface heat is constantly transported to the
night side. Multi-band thermal measurements (single- or multiple-epoch data)
can therefore be used to constrain the orientation of the object at
the time of the observations. Here, the knowledge of the (approximate)
rotation period is important for such investigations. In the case of our
3-band VISIR data we find that the observed spectral N-band slope is
much better explained by temperatures connected to an equator-on viewing
geometry. A pole-on temperature distribution would produce systematically
higher J8.9 fluxes and lower PAH2\_2, that is, a less-steep SED slope in
the N-band wavelength range. These equator-on geometries correspond to
a rotation axis pointing towards ($\lambda_{ecl}$, $\beta_{ecl}$) =
(67$^{\circ}$, +71$^{\circ}$) for a prograde rotation and
(67$^{\circ}$, -71$^{\circ}$) for a retrograde rotation ($\pm$ $\approx$30$^{\circ}$).

A more accurate test would require thermal measurements covering a
wider wavelength range. The VISIR instrument is equipped with an M-band
filter (bandpass 4.54 - 5.13\,$\mu$m), but no M-band measurements were included in our
programme. In this context, we also searched for possible
NEOWise (Mainzer et al.\ \cite{mainzer14}) detections
at shorter wavelengths (3.4 and 4.6\,$\mu$m), but our target was either
too faint (Jun.\ and Nov.\ 2014, Aug.\ 2015, Jun.\ 2016) or too bright
(Oct.\ 2015). In Oct.\ 2015, the Halloween asteroid was very close to Earth
and it crossed the solar elongation zone visible by NEOWise (approximately
87$^{\circ}$ - 93$^{\circ}$) with an apparent motion of more than
10$^{\prime \prime}$/s. At the same time, the object was extremely bright
with an estimated W2 (4.6\,$\mu$m) flux of more than 10\,Jy, well above
the NEOWise saturation limits. We are not aware of any other auxiliary
thermal measurements of 2015~TB$_{145}$. 

\subsection{Thermal inertia}

If we look at an object pole-on, the temperature distribution
does not depend significantly on the thermal inertia of the surface, but
it does have a significant effect on viewing geometries close to equator-on:
a large thermal inertia transports more engery to the night side
than a low thermal inertia, and the sense of rotation determines
if we see a warm or cold terminator. In our case, we observed the
object at approximately 34$^{\circ}$ phase angle. The prograde rotation
requires a relatively large thermal inertia close to 1000\,Jm$^{-2}$s$^{-0.5}$K$^{-1}$
to explain our measurements, while a retrograde rotation would lead
to $\Gamma$-values close to 400\,Jm$^{-2}$s$^{-0.5}$K$^{-1}$. It is
not possible to distinguish these two cases from our limited observations.
These two minima also depend on the object's rotation period: taking
the second-best period (approximately 5\,h) would shift both $\Gamma$-minima
to larger values at approximately 500\,Jm$^{-2}$s$^{-0.5}$K$^{-1}$ (retrograde)
and 1250\,Jm$^{-2}$s$^{-0.5}$K$^{-1}$ (prograde). A similar effect can
be seen when changing the surface-roughness settings in the TPM: rougher
surfaces lead to larger thermal inertias in the radiometric analysis.
This degeneracy cannot be completely broken by our limited observations.
One would need more thermal data from different phase angles (before and after
opposition) combined with data from a wider wavelength range to
be able to distinguish between roughness and thermal inertia effects.
Here in our case, the fit to the measurements is better when assuming
low-roughness surfaces (r.m.s.\ of surface slopes below 0.5) with
$\chi^2$ minima close to 1.0. Very high-roughness cases, similar to the pole-on
geometries, lead to poor fits to our measurements and the corresponding
$\chi^2$ minima are far from 1.0.
Overall, NEAs tend to have thermal inertias well below 1000\,Jm$^{-2}$s$^{-0.5}$K$^{-1}$
(Delbo et al.\ \cite{delbo15}). If we accept this as a general (natural?) limit,
then we have to conclude that 2015~TB$_{145}$ is very likely to have a retrograde rotation and
that the most-likely thermal inertia is approximately 400-500\,Jm$^{-2}$s$^{-0.5}$K$^{-1}$.
Based on our limited coverage of the object's SED, and unkowns in rotation
period, surface roughness, and possibly also the presence of shallow spectral features, the
derived inertias have considerable uncertainties. For our retrograde solution
we find that the thermal measurements are compatible with $\Gamma$-values in the
range 250 to 700\,Jm$^{-2}$s$^{-0.5}$K$^{-1}$.

\subsection{Size}

Radiometric techniques are known to produce highly-reliable size estimates
(e.g. M\"uller et al.\ \cite{mueller14}, and references therein). But in
cases with limited observational data from only one apparition, the situation
is less favorable. Unknowns in the object's rotational properties lead directly
to large error bars for the derived thermal inertias. For our retrograde,
equator-on situation we find thermal inertias from 250 to 700\,Jm$^{-2}$s$^{-0.5}$K$^{-1}$
for a best-fit to the thermal slope. The translation into radiometric sizes
produces values between 630\,m (low $\Gamma$) and 695\,m (high $\Gamma$). A
rougher surface would also require higher thermal inertias, but the resulting
sizes would be very similar. Also, a longer rotation period pushes the acceptable
inertias to slightly larger values, but roughly equal sizes. First indications
from radar measurements\footnote{Goldstone Radar Observations Planning:
2009~FD and 2015~TB$_{145}$. NASA/JPL Asteroid Radar Research. Retrieved
2015-10-22: {\tt http://echo.jpl.nasa.gov/\-asteroids/\-2009FD/\-2009FD\_planning.html}}
gave a size estimate of approximately 600\,m. Our analysis
shows that even extreme settings (low roughness, fast rotation and perfect equator-on
geometry) would require thermal inertias below 150\,Jm$^{-2}$s$^{-0.5}$K$^{-1}$
to produce a radiometric size of 600\,m. The corresponding reduced $\chi^2$ value
in Fig.~\ref{fig:chi2} is above 1.5 and the fit (especially to the J8.9 data) is
very poor. However, we cannot completely rule out the 600-m size (in combination with
a thermal inertia of approximately 100-200\,Jm$^{-2}$s$^{-0.5}$K$^{-1}$). One explanation could
be that fine-grained regolith material affects the surface emissivity in
a strongly wavelength-dependent way (while we assume a flat emissivity of 0.9),
with approximately 10\% higher emissivities in the SIV\_2 band compared to the J8.9 band.
Studying the emissivity variations
found for three Trojans (Emery et al.\ \cite{emery06}), however, we believe that such
strong variations in the 9-12\,$\mu$m range are not possible. The poor $\chi^2$ fit
of a 600-m body (with or without spectral emission features in the N-band) make
such a size-solution very unlikely.
Taking all these
aspects into account, we estimate a minimum size of approximately 625\,m, and a maximum
of just below 700\,m for the NEA 2015~TB$_{145}$. Our best-fit value is 667\,m
($\Gamma$ = 400\,Jm$^{-2}$s$^{-0.5}$K$^{-1}$, retrograde rotation) or 644\,m
($\Gamma$ = 900\,Jm$^{-2}$s$^{-0.5}$K$^{-1}$, prograde rotation).

\begin{figure}[h!tb]
 \rotatebox{0}{\resizebox{\hsize}{!}{\includegraphics{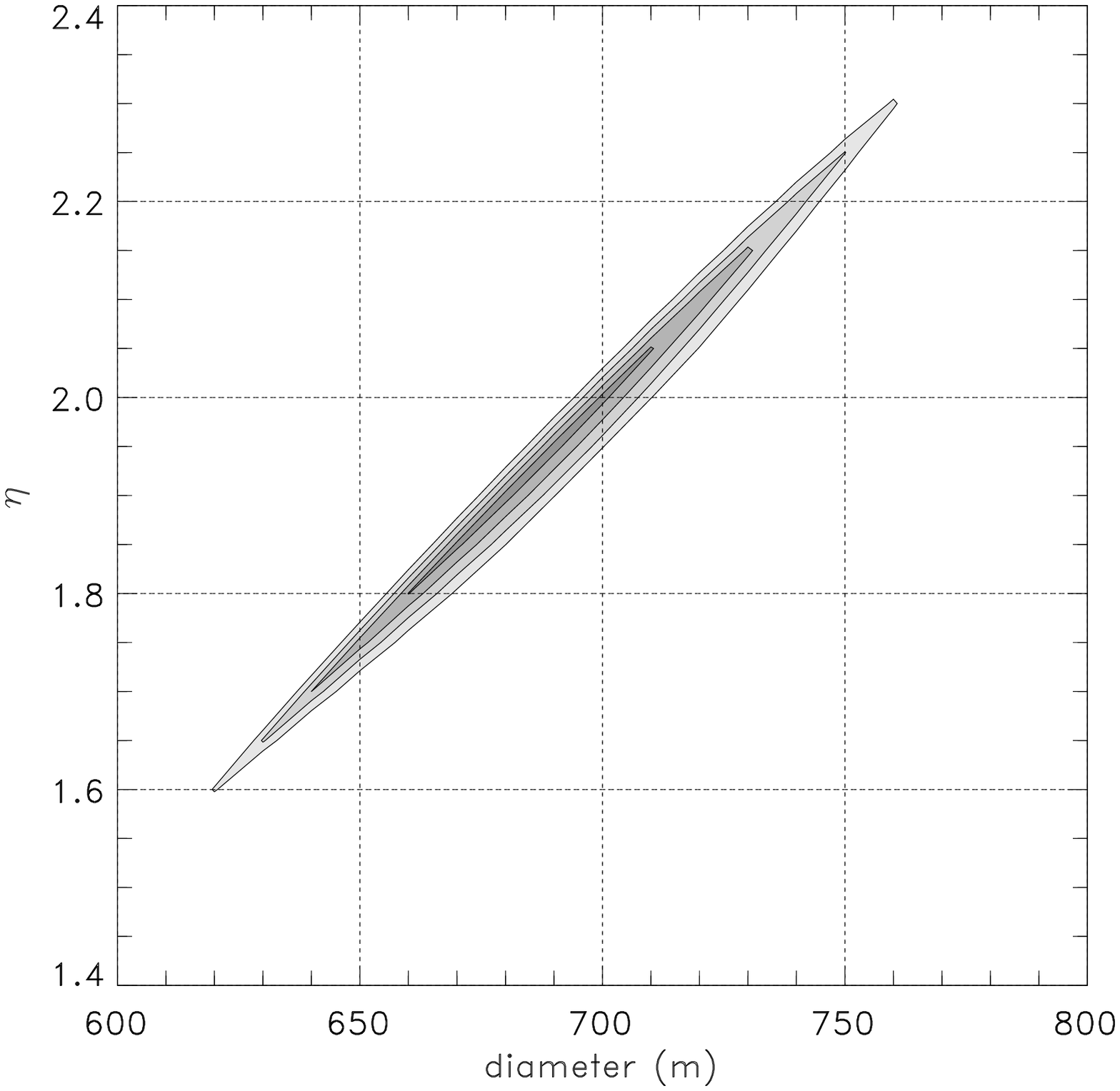}}}
 \rotatebox{0}{\resizebox{\hsize}{!}{\includegraphics{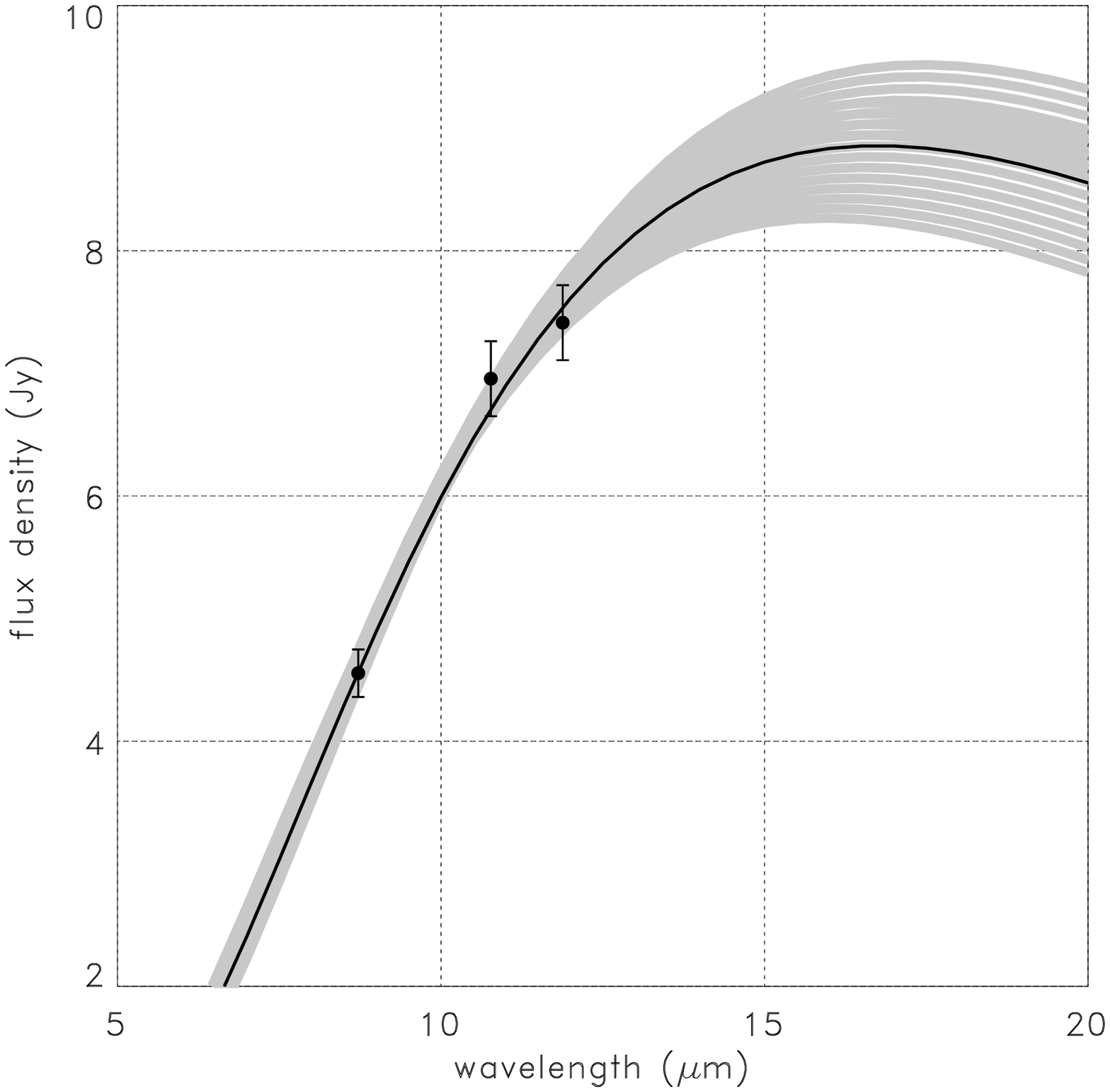}}}
  \caption{NEATM analysis of averaged 3-band VISIR data. Contours in the diameter-$\eta$ plot (top)
           correspond to reduced $\chi^2$ values of  1.0, 1.3, 1.6, and 1.9. The best-fit
           solution (bottom) is shown as a solid black line. Grey curves have $\chi^2_{red}$ 
           below 1.9.
     \label{fig:neatm}}
\end{figure}

An independent NEATM approach (Harris \cite{harris98}) in analysing the VISIR
measurements confirms our large size for 2015~TB$_{145}$ which exceeds the radar
estimates. Figure~\ref{fig:neatm} shows the results in terms of size as a function
of the beaming parameter $\eta$ (top) and the fit to our averaged 3-band fluxes
connected to the observing geometry at half way through the VISIR measurements (see
Tables~\ref{tbl:obsgeometry} and \ref{tbl:obsvisir}). For these calculations we
assumed a spherical shape and a wide range for $\eta$. We determined the radiometric
NEATM size for each $\eta$ together with the corresponding reduced $\chi^2$ value.
The best fit is found for a size of D = 690\,m, $\eta$ = 1.95 and a geometric V-band
albedo p$_V$ = 0.05 (assuming H$_V$ = 19.75\,mag). 

The optimum $\eta$ is relatively high, but justified considering the fast
rotation (2.9\,h) of a kilometer-sized object.
There is a strong correlation between the possible diameters and $\eta$ values
and a simple error estimate is not easy, but following the previous TPM analysis,
the object's size is found to be significantly larger than 600\,m.

\subsection{Albedo}

Calculation of the geometric albedo requires a robust estimate of the
object's absolute magnitude. Here again, the H-magnitude calculations suffer
from low-quality single-apparition measurements. Our case with observations
from an extremely restricted phase angle range is probably not typical, but often the
H-mag calculations suffer from unknowns in the phase relation. Our own calibrated
R-band measurements are all taken close to 34$^{\circ}$ phase angle and the MPC
R-/V-band database entries were optimised for astrometric calculations and have
no error bars. Simply plotting the reduced magnitudes from MPC shows scatter on
a 1-2 magnitude scale within a given night, bearing in mind that the object's lightcurve
amplitude is only approximately 0.1\,mag. We applied different techniques to derive
an absolute magnitude: the two methods, standard H-G and H-G$_{12}$ , produce
H$_V$ solutions of 19.3\,mag and 19.75\,mag, respectively. The true uncertainty
could even be in the order of 1-2\,mag, depending on the different phase functions
and fitting routines (non-linear vs. linear). Such large uncertainties are 
probably common among the single-apparition NEAs. In the absence of thermal
measurements, these H-mag estimates are commonly used to derive sizes for
low- and high-albedo assumptions. Here, a 19\,mag value for a low-albedo
(p$_V$ = 0.03) object leads to a size estimate above 1.2\,km, while 20\,mag combined
with p$_V$ = 0.5 would result in a size estimate of 188\,m, smaller in size by a
factor of over 6. Adding thermal data changes the situation dramatically
and produces reliable size-albedo solutions. However, using a H-mag of 19.0
pushes the geometric albedo to approximately 10\%, while an absolute magnitude of
20.0 would give a 4\% albedo. This uncertainty in albedo can only be
reduced with more reliable photometric data points spread over a wider
phase angle range.

\subsection{Shape}

Both the visual lightcurves and the thermal measurements show 
brightness/flux variations over timescales of a few hours. After correcting for
the rapidly changing observing geometries, we find lightcurve amplitudes
close to 10\%. This is compatible with a rotating ellipsoid with
an axis ratio a/b = 1.1 seen equator-on. More elongated bodies are
also possible, but then in combination with a different obliquity for
the spin axis. An extreme axis ratio of a very elongated object
can be excluded with high probability, mainly because of the observed
thermal emission spectrum which can be best explained by equator-on
viewing geometries. On the other hand, a regular ellipsoidal shape can
also be excluded by the strong deviations of the lightcurve from a
sinosoidal shape (see Figs.~\ref{fig:best_period_lc},
\ref{fig:secondbest_period_lc}, and \ref{fig:composite}).

\section{Conclusions}
\label{sec:con}

NEAs with very close encounters with Earth always receive great attention
from the public. It is usually also relatively easy to obtain high-S/N photometry
during the encounter phase, even with small telescopes and for very small
objects. However, a reliable and high-quality physical and thermal
characterisation of the objects with single-apparition data is often
challenging. The Halloween asteroid had an encounter with Earth at
1.3 lunar distances on October 31, 2015, and nicely
illustrates the possibilities and limitations of current analysis techniques.

The calculation of a reliable H-magnitude is difficult and depends strongly on
                 a good photometric coverage over wide phase angle ranges. For 2015~TB$_{145}$
                 there are no measurements available at phase angles below 33$^{\circ}$ and we
                 obtained H$_V$ magnitudes between 19 and 20\,mag, depending on the applied
                 phase relation. Using the H-G$_{12}$ conventions for sparse and low-accuracy
                 photometric data (Muinonen et al.\ \cite{muinonen10}), we find H$_R$ = 19.2\,mag
                 and H$_V$ = 19.8\,mag, with large uncertainties which could easily be up to one
                 magnitude (conservative estimate based on fitting various phase functions
                 (H-G, H-G$_{12}$, H-G$_1$-G$_2$) and taking the whole range of all possible
                 solutions and uncertainties into account).
The determination of the object's V-R colour is more reliable and based on
                 different calculations and different data sets, we find a V-R colour of
                 0.56 $\pm$ 0.05\,mag.

We combined lightcurve observations from different observers with our own
                 measurements. The Fourier analysis periodogram shows several possible
                 rotation periods, with two periods producing best
                 $\chi^2$ minima and an acceptable fit of various lightcurve fragments 
                 relative to each other: 2.939 $\pm$ 0.005\,hours (amplitude 0.12 $\pm$ 0.02\,mag)
                 and 4.779 $\pm$ 0.012\,h (amplitude 0.10 $\pm$ 0.02\,mag).
A similar lightcurve amplitude is also seen in the thermal measurements
                 (after correcting for the rapidly changing Earth-NEA distance), but 
                 measurement errors are too large to constrain the rotation period.

The detemination of the object's spin axis orientation is not possible
                 from such sets of lightcurves. 2015~TB$_{145}$ was observable only for
                 approximately two weeks in October 2015 and neither the phase angle bisector nor the 
                 phase angle changed significantly and no noticeable change in rotation period or amplitude
                 was seen (Warner et al.\ \cite{warner16}) (also confirmed by our additional
                 lightcurve data).

For our TPM radiometric analysis we find 2015~TB$_{145}$ was very likely
                 close to an equator-on observing geometry ($\pm$ $\approx$30$^{\circ}$)
                 during the time of the VLT-VISIR measurements. A pole-on geometry can
                 be excluded from the observed 8-12\,$\mu$m emission slope from the
                 multiple 3-filter N-band measurements.
In the process of radiometric size and albedo determination from the
                 combined thermal data set we find the object's thermal inertia
                 to be between 250 to 700\,Jm$^{-2}$s$^{-0.5}$K$^{-1}$ in case of a
                 retrograde rotation, and above $\approx$500\,Jm$^{-2}$s$^{-0.5}$K$^{-1}$
                 for a prograde rotation. These ranges are found for an equator-on observing
                 geometry of a spherical body with 2.939\,h rotation period and a low surface
                 roughness (r.m.s.\ of surface slopes of 0.1). Moving away from an equator-on
                 geometry, or using longer rotation periods or higher surface roughness would
                 shift these thermal inertia ranges to larger values.
The maximum (model) surface temperatures during our VISIR measurements are
                 close to 350\,K, (equator-on geometry) or even above in case of a spin
                 axis closer to a pole-on geometry.
From our radiometric TPM analysis we estimate a minimum size of approximately 625\,m,
                 and a maximum of just below 700\,m for the NEA 2015~TB$_{145}$ (the corresponding
                 NEATM diameter range is between 620 and 760\,m, for beaming parameters $\eta$ of
                 1.6 and 2.2, respectively, and best-fit values at 690\,m for $\eta$ = 1.95).
The best match to all thermal measurements is found for:
                 (i) Thermal inertia $\Gamma$ = 900\,Jm$^{-2}$s$^{-0.5}$K$^{-1}$;
                 D$_{eff}$ = 644\,m, p$_V$ = 5.5\% (prograde rotation with 2.939\,h);
                 (ii) thermal inertia $\Gamma$ = 400\,Jm$^{-2}$s$^{-0.5}$K$^{-1}$;
                 D$_{eff}$ = 667\,m, p$_V$ = 5.1\% (retrograde rotation with 2.939\,h).

The reconstruction of the object's shape is not possible, but an equator-on
                 viewing geometry combined with the visual and thermal lightcurve amplitude
                 would point to an ellipsoidal shape with an axis ratio a/b of approximately
                 1.1 (assuming a rotation around axis $c$).

Following the discussions and formulas in Gundlach \& Blum (\cite{gundlach13})
                 we can also estimate possible grain sizes on the surface of the NEA 2015~TB$_{145}$.
                 For the calculations we used the CM2 meteoritic sample properties (matching our low
                 albedo of approximately 5\%) from Opeil et al.\ (\cite{opeil10}), with a density
                 $\rho$ = 1700 kg\,m$^{-3}$, and a specific heat capacity of the regolith
                 particles $c$ = 500\,J\,kg$^{-1}$\,K$^{-1}$. A thermal inertia of 400\,Jm$^{-2}$s$^{-0.5}$K$^{-1}$
                 leads to heat conductivities $\lambda$ in the range 0.3-1.9\,W\,K$^{-1}$\,m$^{-1}$,
                 or 2-12\,W\,K$^{-1}$\,m$^{-1}$ for a thermal inertia
                 of 1000\,Jm$^{-2}$s$^{-0.5}$K$^{-1}$. The estimated grain sizes would then be
                 in the order of 10-20\,mm ($\Gamma$ = 400\,Jm$^{-2}$s$^{-0.5}$K$^{-1}$) or
                 50-100\,mm ($\Gamma$ = 1000\,Jm$^{-2}$s$^{-0.5}$K$^{-1}$), considering a wide
                 range of regolith volume-filling factors from 0.1 (extremely fluffy) to 0.6
                 (densest packing).

Based on our analysis and careful inspection of the available data, we make several
recommendations for future observations of similar-type objects:
\begin{enumerate}
\item Size estimates for single-apparition NEAs from H-magnitudes alone are highly uncertain.
                 The estimates depend on assumptions for the albedo, but also suffer from possible huge 
                 uncertainties in H-magnitudes which can easily reach 1-2\,mag. Here, without
                 thermal data, one would estimate a size below 200\,m 
                 (H = 20.0, p$_V$ = 0.5) or above 1.2\,km (H = 19.0, p$_V$ = 0.03). Having
                 a more accurate H-magnitude would shrink the possible sizes to values
                 between approximately 250 and 950\,m.
\item For the determination of reliable H-magnitudes it is essential to
                 have calibrated R- or V-band photometric data points spread over a wide range
                 of phase angles, preferably also including small phase angles below 7.5$^{\circ}$.
\item For a good-quality physical characterisation (radiometric analysis and also the
                 interpretation of radar echos) it is important to find the object's rotational
                 properties. This requires high-quality lightcurves: (i) in cases of very short rotation 
                 periods it is important to avoid rotational smearing, with exposures well below $\approx$0.2
                 of the rotation period (Pravec et al.\ \cite{pravec00}); (ii) for rotation periods of several
                 hours or longer there is the challenge of covering substantial parts of the lightcurve
                 of very fast-moving objects with one set of reference stars; rapidly changing image FOVs
                 cause severe problems in reconstructing the composite lightcurve. Thus, observing
                 instruments with large FOVs of the order of one or a few degrees are recommended.
\item For thermal measurements of NEAs it is very helpful to have (i) the widest possible wavelength
                 coverage, possibly also including M- or Q-bands, to constrain the thermal inertia from
                 the reconstructed thermal emission spectrum; (ii) well-calibrated measurements (calibration
                 stars close in flux, airmass, time and location on the sky) to derive reliable size estimates;
                 (iii) observations before and after opposition at different phase angles to determine the
                 sense of rotation, to put strong constraints on thermal inertia and to break the 
                 degeneracy between thermal inertia and surface roughness effects.
\end{enumerate}

The next encounter of 2015~TB$_{145}$ with Earth is in November 2018 at a distance of approximately 0.27\,AU
and with an apparent magnitude of approximately 19.5\,mag.
During that apparition it would easily be possible to get
reliable R-band lightcurves for 2015~TB$_{145}$. The object will reach approximately
5, 60 and 90\,mJy at 5, 10 and 20\,$\mu$m, respectively, detectable with useful S/N ratios with
ground-based MIR instruments such as VLT/VISIR. In the future, with telescope sizes well above
10\,m, one can expect to study Apollo asteroids comparable to 2015~TB$_{145}$ out to
distances of up to 0.5\,AU.

\begin{acknowledgements}
  We would like to thank the ESO staff at Garching and Paranal for
  the great support in preparing, conducting, and analysing these challenging
  observations of such an extremely fast-moving target with very
  short notice, with special thanks to Christian Hummel, Konrad Tristram, and
  Julian Taylor. The research leading to these results has received
  funding from the European Union's Horizon 2020 Research and
  Innovation Programme, under Grant Agreement no 687378.
  We would like to thank Victor Al{\'i}-Lagoa for providing feedback to a 
  potential detection of 2015~TB$_{145}$ by NEOWise.
  This research is partially based on observations collected at Centro Astron{\'o}mico
  Hispano Alem{\'a}n (CAHA) at Calar Alto, operated jointly by the
  Max-Planck Institut f\"ur Astronomie (MPIA) and the Instituto de Astrof{\'i}sica
  de Andaluc{\'i}a (CSIC). This research was also partially based on observation
  carried out at the Observatorio de Sierra Nevada (OSN)
  operated by Instituto de Astrof{\'i}sica de Andaluc{\'i}a (CSIC). Funding
  from Spanish grant AYA-2014-56637-C2-1-P is acknowledged. 
  Hungarian funding from the NKFIH grant GINOP-2.3.2-15-2016-00003 is also acknowledged.
  R.\ D.\ acknowledges the support of MINECO for his Ramon y Cajal Contract. 
\end{acknowledgements}

\end{document}